\documentclass[a4paper,fleqn,usenatbib]{mnras}

\usepackage{newtxtext,newtxmath}
\usepackage[T1]{fontenc}
\usepackage{ae,aecompl}
\usepackage{graphicx}	
\usepackage{amsmath}	
\usepackage{amssymb}

\title[Unique partial eclipse of IGR J16393-4643]{Investigating a unique partial eclipse in the High Mass X-ray Binary 
IGR J16393-4643 with {\it Swift}-XRT}

\author[S. Kabiraj, N. Islam and B. Paul]{Sanhita Kabiraj$^{1,2}$\thanks{Contact e-mail: \href{mailto:sanhita@rri.res.in, 
sanhita@iisc.ac.in}
{sanhita@rri.res.in, sanhita@iisc.ac.in}}
Nazma Islam$^{3}$ and 
Biswajit Paul$^{1}$\\
$^{1}$Raman Research Institute, Sadashivnagar, Bangalore-560080, India\\
$^{2}$Joint Astronomy Programme, Indian Institute of Science, Bangalore-560012, India\\
$^{3}$ Center for Astrophysics | Harvard \& Smithsonian Center for Astrophysics, 60 Garden Street, Cambridge, MA 02138, USA}

\date{Accepted XXX. Received YYY; in original form ZZZ}

\pubyear{2018}

\begin{document}
\label{firstpage}
\pagerange{\pageref{firstpage}--\pageref{lastpage}}
\maketitle

% Abstract of the paper
\begin{abstract}
The orbital profile of the High Mass X-ray binary IGR J16393-4643 shows a dip in its X-ray intensity, which was previously interpreted as an eclipse. Unlike most eclipsing HMXBs, where the X-ray eclipses are about two orders of magnitude fainter compared to the out of eclipse emission, this particular eclipse like feature is narrow and partial, casting doubt if it is indeed an eclipse. To further investigate the nature of this low intensity orbital phase, we use a large number of observations with {\it{Swift}}-XRT, covering the entire orbital phase. The soft X-ray observations also show this low intensity phase, which is about 30\% of the intensity during rest of the orbit. We also carried out orbital phase resolved spectroscopy to compare the change in the spectral parameters inside and outside of this low intensity state. The results indicate that this low intensity state might not be an eclipse, as previously thought but absorption in the stellar corona. We have also provided the inclination angle of the binary for grazing eclipse caused by the stellar corona.

 \end{abstract}


%%%%%%%%%%%%%%%%%%%%%%%%%%%%%%%%%%%%%%%%%%%%%%%%%%

%%%%%%%%%%%%%%%%% BODY OF PAPER %%%%%%%%%%%%%%%%%%-ray

\section{Introduction}
In an X-ray binary system, the X-ray emitting compact star can get occulted by its companion in every orbit and the X-ray intensity observed from the binary is reduced by a large factor. Variation in the orbital intensity depends upon the inclination of the orbital plane of the X-ray binary with respect to our line of sight, and angular size of the companion star to the compact star. For compact emission regions, the eclipse is expected to be complete, except any scattered radiation from the environment. For Low Mass X-ray Binaries (LMXBs), eclipses are predicted for a narrow range of inclination as the companion star usually has a smaller angular diameter to the compact star \citep{1987A&A...178..137F}. The only exception to this can be the Accretion Disk Corona (ADC) sources among the LMXBs like 4U 1822-37, which shows a partial eclipse as the ADC is comparable to the size of the companion star \citep{2010MNRAS.409..755J}. Accretion disk corona was first used to explain narrow partial eclipses \citep{1982_White_Holt}. The X-rays from the compact object heats up the gas in the corona which gives rise to residual flux which can be observed even during the eclipses. The presence of a corona can cause scattering of photons which leads to broad and shallow dips in the X-ray light curve. We also see low X-ray intensity states prior to the eclipse in some High Mass X-ray binaries (HMXBs) due to the presence of enhanced stellar wind in certain orbital phases, which lead to an orbital phase dependent absorption.

\begin{figure*}
\centering
\includegraphics[scale=0.45, angle=-90]{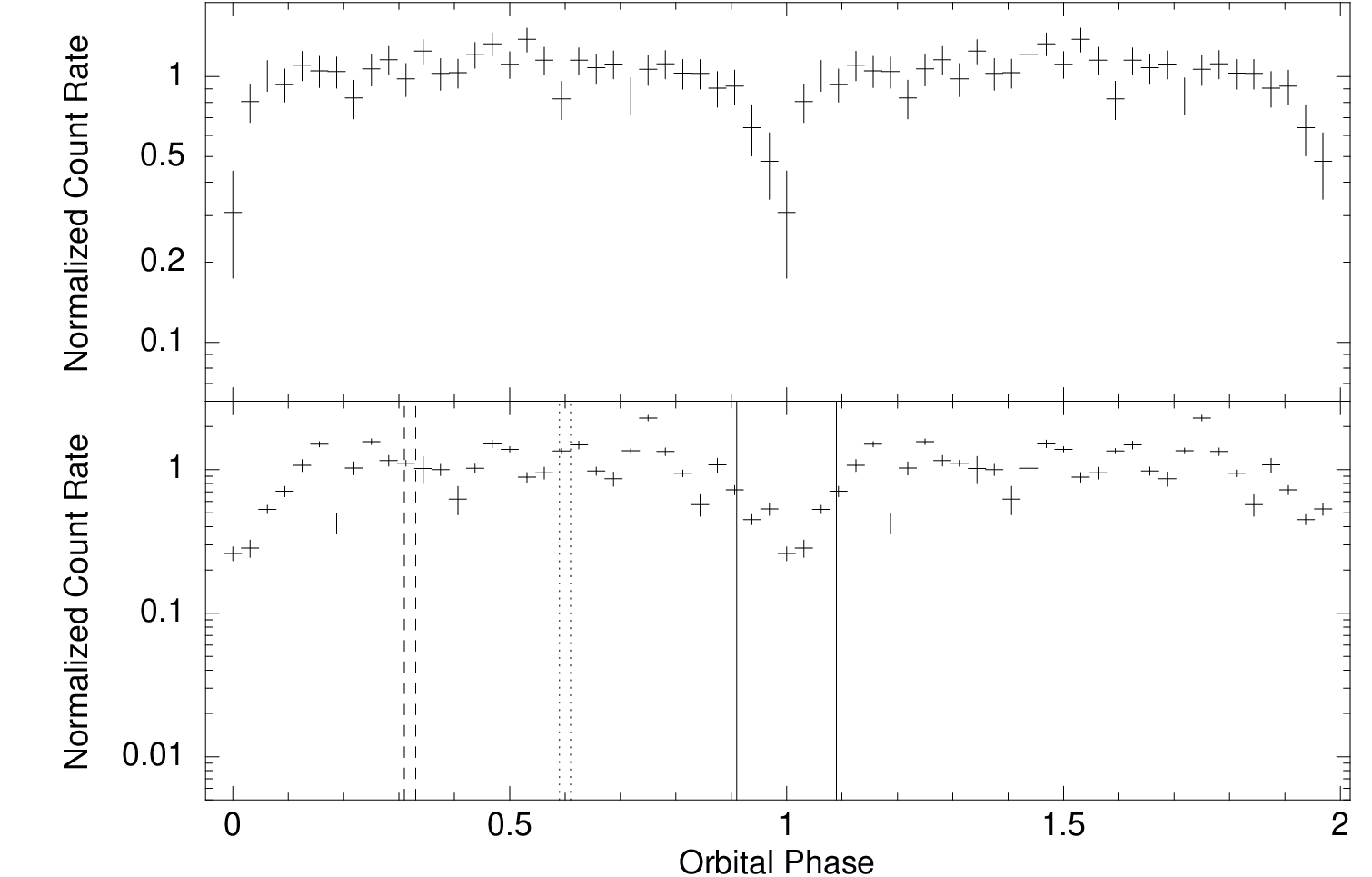} 
\caption{Top panel: The intensity profile of HMXB IGR J16393-4643 constructed with {\it Swift}-BAT
lightcurve folded with a period of 366106 seconds with the epoch of MJD 53418.300 to have the minimum counts at phase 1. The low intensity state appears to be an eclipse. \newline
Bottom panel: \textit{Swift}-XRT light curves, folded with the same period. The dashed and dotted lines enclose the
phase 0.31-0.33 and 0.59-0.61 respectively and the region between the two solid lines (i.e. phase 0.91-1.09) indicates
the partial eclipse phase.}
\label{BAT&XRT}
\end{figure*}

\par 
 In an HMXB system with a neutron star, the X-ray emission region sizes about a few kilometers, whereas the typical velocity of a neutron star in such system is of the order of 200 km/s and the companion star is  many orders of magnitude larger in size compared to the X-ray emission region. Therefore, X-ray eclipses are expected to be sharp and complete \citep{2015A&A...577A.130F}. The eclipse ingress and egress are however gradual in soft X-rays due to strong absorption in the stellar wind which has large column density during the eclipse ingress and egress. However, hard X-ray eclipses are always known to be sharp. IGR J16393-4643 is an unusual HMXB system with a neutron star that shows a gradual and partial eclipse even in the hard X-rays \citep{2015MNRAS.446.4148I}.
\par 
IGR J16393-4643 was detected in the INTEGRAL survey of the galactic plane and was identified to be the same source as AX J1639.0-4642, discovered earlier with ASCA observatory \citep{2004ApJ...607L..33B, 2001ApJS..134...77S}. It belongs to the highly absorbed galactic HMXBs and is an X-ray pulsar with a spin period of $\sim 910$ seconds \citep{Bodaghee2006}.
%\textcolor{red}{Change the previous sentenc to (no need to give the coordinates): It belongs to the family of persistent and heavily photo-absorbed galactic HMXBs and is an X-ray pulsar with a spin period of $\sim 910$ seconds.}
The spectral characteristics of IGR J16393-4643 is indicative of a heavily absorbed wind accreting HMXB with OB type supergiant counterpart \citep{Bodaghee2006, 2004_Walter}. 
%\textcolor{red}{Bodaghee et al. reported photon index, not spectral index. The photon index in hard X-rays was found to be high, but it was just about 1.0 in XMM band and in the same paper the author clarified that the large spectral index in ISGRI band was due to a cutoff and E-fold. So don't mention this high spectral index in the previous sentence. Anyway why would high spectral index indicate a BH type supergiant?}
Also, the XMM and NuSTAR spectra detected the presence of iron emission lines at 6.4 keV and 7.1 keV and cyclotron line at 29.3 keV, indicating a strong magnetic field of about 10$^{12}$ Gauss \citep{Bodaghee2006,2016ApJ...823..146B}. \citet{Bodaghee2012} suggested that the counterpart of the pulsar is possibly a distant reddened star blended with a bright 2MASS star. Under this assumption, it is either a supergiant O9 star with R$\sim$20R$_{\odot}$ or a main sequence B star with R$\sim$10R$_{\odot}$. For an O9 star, the estimated distance of 25 kpc is too far to be feasible, but for a main sequence B star, the estimated distance is 12 kpc which is physically reasonable due to the presence of a star-forming H II region at the same distance \citep{2003_Russeil}.
\par 
%\textcolor{red}{RXTE lightcurves did not show the presence of eclipse. Please correct the statement.}
The X-ray lightcurve from Swift-BAT showed a periodicity of $\sim 4.24$ days with an apparent eclipse with an eclipse semi-angle $\sim 17 \degr$ \citep{2014AAS...22332308C, 2015MNRAS.446.4148I}. From the eclipse duration, \citet{2015MNRAS.446.4148I} determined the orbital inclination to be $39\degr$-$57\degr$ for a companion with radius $R\sim20R_\odot$ and $60\degr$-$77\degr$ for a companion with radius $R\sim10R_\odot$. Interestingly, the eclipse profile of IGR J16393-4643 demonstrates a partial eclipse of the source in the hard X-ray energy band, unlike other eclipsing HMXBs. There is approximately 25\% amplitude during the eclipse phase in comparison to the mean flux. The eclipsing HMXBs, show some X-ray emission in their eclipse phase because of X-ray reprocessing from stellar wind or accretion disk but that is limited within a factor of 1/8 to 1/237 in comparison to the out-of-eclipse flux \citep{2019_Aftab}.
%This intrigued us about the nature of absorption and scattering mechanism in this particular binary pulsar to show a partial eclipse. During the eclipse phase observation from {\textbf{\it{Swift}}} was only available.
\par
In this work, we have taken all {\it Swift}-XRT observations of IGR J16393-4643 which were taken within 2 arcmin radius centering the source having exposure time more than 200 seconds. We have studied orbital intensity profile in the low energy band using a large number of XRT observations. We have carried out X-ray spectroscopy to estimate the spectral parameters, especially the absorption column density and the upper limits of iron emission line in different orbital phases by combining multiple observations. We have also analysed the long-term {\it{Swift}-BAT} lightcurve from MJD 53421 to MJD 58173 to compare the orbital intensity profiles in soft (XRT) and hard (BAT) X-rays. Our results can put some useful insight into this unique partial eclipse of IGR J16393-4643.

\begin{table*}
\caption{ Log of all observations with exposure time,  total number of source photons, average count-rate and orbital phase.}
\label{table1}
  \centering
 \begin{tabular}{|c|c|c|c|c|c|c|c|}
 \hline
 &&&&&&\\
 
 Order       &Observation  &Observation   &Exposure    &Total no.     & Average          &Orbital\\
  No.          &MJD          &ID 	  &time        & of photons   &count-rate        &Phase\\
               &             &            &            &              &                  &\\
               &             &            &(seconds)   &(counts)      &(counts/sec)      &\\
          
 \hline
             &            &            &            &                  &                   &\\
  1. &57411 &00034135004 &1056 &117 &0.11 &0.31\\
  2.         &57413       &00034135006 &        939 & 	 89            &	0.09       &0.81\\
  3.         &57417       &00034135010 &        966 & 	118            &	0.12       &0.74\\
  4. &57418 & 00034135011 & 929 & 57 & 0.06 &0.93\\
  5. &57420 &00034135013 &1011 &134 &0.13 &0.59\\
  6.         &57422       &00034135015 &	  1785 &	115 &   0.06	       &0.98\\
  7.         &57423      &00034135016 &	   986 &	252 &   0.25	       &0.15\\
  8.         &57424      &00034135017 &	   864 &	 44 &   0.05	       &0.56\\
  9.         &57425      &00034135018 &	  1064 &	114 &   0.11	       &0.73\\ 
  10. &57426 &00034135019 &999 &15 & 0.02 &0.01\\
  
  11.        &57427      &00034135020 &	   852 &	 91 &	0.11	       &0.23\\
  12.  &57428 &00034135021 &941 &111  & 0.12 &0.31\\
  13.        &57430      &00034135023 &       1321 &	168 &	0.13	       &0.78\\
  14. &57431      &00034135024 &996 &34 &0.03 &0.07\\
  
  15.        &57432      &00034135025 &	   779 & 	 81 &	0.10	       &0.24\\
  16.        &57433      &00034135026&	   847 &	 75 &   0.09	       &0.52\\
  17.        &57434      &00034135027 &	   864 &	101 &   0.12	       &0.78\\
  
  18. &57435 &00034135028 &859 &93 &0.11 &0.12\\
  19.   &57437      &00034135030 &1069 &158 &0.15  &0.59\\
  20.        &57438      &00034135031&	  1046 &         82 &	0.08	       &0.82\\
  21. &57439 &00034135032 &1064 &42 &0.04 &0.06\\
  22.        &57441      &00034135034&        1016 &	121 &	0.12	       &0.51\\
  23.   &57442 &00034135035 &1056 &121 &0.11 &0.60\\
  24.&57443 &00034135036 &859 &42 &0.05 &0.98\\
  
  25.        &57444      &00034135037&	   924 &	 78 &   0.08	       &0.15\\
  26.        &57445      &00034135038&	   217 &	 22 &   0.10	       &0.34\\
  27.        &57446      &00034135039&	  1346 &	150 &	0.11	       &0.61\\
  28.        &57448      &00034135041&	   946 &	 98 &	0.10	       &0.14\\
  29.        &57458      &00034135051&	  1054 &	 95 &   0.09	       &0.45\\
  30.        &57459      &00034135052&	   969 &	 91 &   0.09	       &0.65\\
  
  31.        &57462      &00034135055&	   984 &	169 &	0.17	       &0.50\\
  32.        &57464      &00034135057&	   946 &	 32 &   0.03	       &0.92\\

  33.        &57466      &00034135059&	  1021 &	101 &	0.10	       & 0.33\\ 
  %       &57468      &00034135061&	    45 &	  3 &   0.07	       &0.68\\
  34. &57469 &00034135062 &817 &75 &0.09 &0.10\\

  %57470      &00034135063&	    42 & 	  3 & 	0.07	       &0.19\\
  35.        &57471      &00034135064&	   330 &	 55 &	0.17	       &0.63\\
  %57472      &00034135065&	    45 &	  4 &	0.09	       &0.84\\
  36. &57473 &00034135066 & 789 &25 &0.03 & 0.07\\
  37.        &57479      &00034135072&	  1039 &	107 &	0.10	       &0.50\\

  38.&57481 &00034135074 &916 &70 &0.08 &0.95\\
  39.        &57483      &00034135076&	  1031 &	119 &	0.12	       &0.44\\   
  40.        &57484      &00034135077&	   944 &	 90 &	0.10	       &0.72\\
  41.        &57485      &00034135078&	   897 &	352 &	0.39	       &0.74\\
  42.&57486 &00034135079 &944 & 38 &0.04 &0.98\\  
  43.&57490 &00034135083 &844 &59 &0.07  & 0.07\\
  44.        &57491      &00034135084&	   986 &	152 &	0.15	       &0.26\\
  45. &57492 &00034135085 &954 &141 &0.15 &0.60\\
  46.        &57493      &00034135086&	   971 &	148 &	0.15	       &0.62\\
  47.        &57496      &00034135089&	  1074 &	179 &	0.17	       &0.50\\
  48.        &57497      &00034135090&	   976 &	103 &	0.11	       &0.78\\
  49.        &57498      &00034135091&	   912 &	70 &	0.08	       &0.89\\
  50.&57499 &00034135092 &594 &24 &0.04 &0.06\\
  51.&57503 &00034135096 &1091 &84 &0.08 &0.07\\
  52.        &57504      &00034135097&	   996 &	96 &	0.10	       &0.38\\
  53.        &57506      &00034135099&	  1036 &	78 &	0.08	       & 0.83\\
  54.        &57508      &00034135101&	  1044 &       122 &	0.12	       &0.29\\
  55.        &57509      &00034135102&	  1039 &	93 &	0.09	       &0.53\\

    &&&&&&\\
 \hline
 \end{tabular}
 \end{table*}
 
 \begin{table*}
   \centering
 \begin{tabular}{|c|c|c|c|c|c|c|c|}
 \hline
 &&&&&&\\
 
  Order        &Observation  &Observation   &Exposure    &Total no.     & Average          &Orbital\\
  No.          &MJD          &ID 	    &time        & of photons   &count-rate        &Phase\\
               &             &              &            &              &                  &\\
               &             &              &(seconds)   &(counts)      &(counts/sec)      &\\

 \hline
             &           &            &            &        &                  &  \\
  56.        &57510      &00034135103&	   922 &	162 &	0.18	       &0.79\\  
  57.        &57514      &00034135107&	   894 &	126 &	0.14	       &0.64\\
  58.        &57516      &00034135109&	   917 &	 39 &	0.04	       &0.19\\
  59.	     &57519      &00034135112	&  587 &         27 &   0.05           &0.91\\
  60.        &57524      &00034135117   &1044 & 32 & 0.03 &0.01\\
  61.       &57525      &00034135118    &922 &100 &0.11 &0.32\\
  62.       &57527       &00034135120   &211 &96 & 0.45 &0.79\\
  63.        &57528      &00034135121&	  1126 &	119 &	0.11	       &0.89\\
  64.        &57530      &00034135123&	   862 &	 47 &	0.05	       &0.47\\
  65.        &57531      &00034135124&	   912 &	 80 &	0.09	       &0.69\\
  66.&57532 & 00034135125 &979 & 18 & 0.02 & 0.02\\
  67.        &57533      &00034135126&	   862 &	194 &	0.23	       &0.25\\
  68. &57538 & 00034135131 & 989 &107 &0.11 &0.32\\
  69.&57549 &00034135142 &954  & 26 & 0.03 & 0.01\\
  70.        &57551      &00034135144&	   969 &	231 &	0.24	       &0.47\\
  71.        &57552      &00034135145&	  1031 &	189 &	0.18	       &0.70\\
  72.        &57553      &00034135146&	  1036 &	 37 &	0.04	       &0.92\\
  73.        &57554      &00034135147&	  1019 &	152 &	0.15	       &0.14\\

   &&&&&&\\
 \hline
 \end{tabular}
 \end{table*}

\section{Observations and Data Analysis}
The Neil Gehrels Swift Observatory is a multi-wavelength mission with three instruments: A) the Burst Alert Telescope (BAT; \citealt{Barthelmy2005}), having an energy range of 15-150 keV; B) the X-Ray Telescope (XRT; \citealt{Burrows2005}), having an energy range of 0.2-10 keV, and the Ultra Violet Optical Telescope (UVOT; \citealt{Roming2005}). The XRT has three readout modes: 1)Imaging Mode (IM), 2)Windowed Timing Mode (WT) and 3)Photon Counting (PC) Mode. Photon Counting mode has a time resolution of 2.5 seconds. We have used the data collected in this mode as it provides full imaging and spectroscopic resolution.\\

We have analyzed 73 observations in 0.2-10 keV {\it Swift}-XRT energy band using HEASOFT 6.22. We have considered all observations of IGR J16393-4643 which were taken within 2 arcmin radius centering the source having exposure time more than 200 seconds. These were taken at different orbital phases from MJD 57411 to MJD 57554. The observation ID, exposure time, photon counts and orbital phase of all of them are tabulated in Table \ref{table1}. For all these observations, we have used cleaned data in photon counting mode. Minimum exposure time among these observations was 211 seconds and the maximum time is 1785 seconds, and the lowest and highest count-rate was 0.02 count/s and 0.39 count/s. The source photons were extracted from a region of 60$''$ radius centering the source, and also the background photons were extracted from a source free region with 60$''$ radius in the FoV, where no other X-ray source was present. We have generated lightcurves and spectra from source and background region files using XSELECT v2.4d.\\

The exposure maps were generated and had been used to create the ancillary response file (ARF). We got the required response matrix file (RMF) from the calibration database CALDB 1.0.2. We have generated the spectra for source and background for
each observation and used the ARF and RMF files to get the required spectra for analysis. Therefore, in total we had 73 lightcurves and spectra to
do the timing and spectral analysis. \\

\begin{figure*}
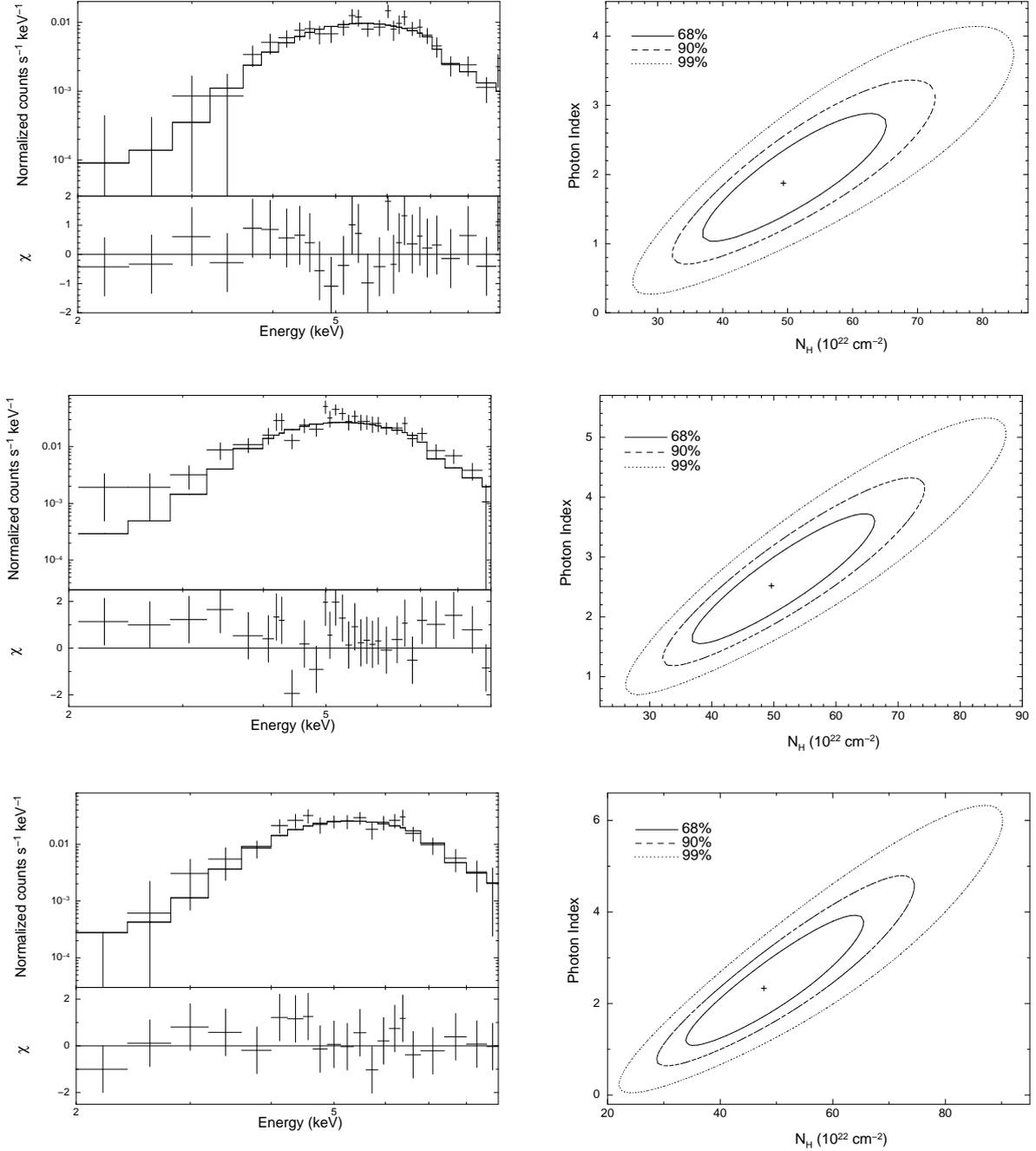

\centering
\vspace*{0.6 cm}
\includegraphics[scale=0.30, angle=-90]{inside_8thMarch2018_powphab.eps} 
\vspace*{0.6cm}
\hspace{5mm}
\includegraphics[scale=0.30, angle=-90]{22aug19_inside_contour.eps}
\vspace*{0.6cm}
\includegraphics[scale=0.30, angle=-90]{28_8thMarch2018_powphab.eps} 
\hspace{5mm}
\includegraphics[scale=0.30, angle=-90]{22aug19_28_contour.eps}
\vspace*{0.6cm}
\includegraphics[scale=0.30, angle=-90]{56_8thMarch2018_powphab.eps} 
\hspace{5mm}
\includegraphics[scale=0.30, angle=-90]{22aug19_56_contour.eps}
\caption{Left column: The added X-ray spectrum from the observations during the partial eclipse phase (at the top),
during phase 0.31-0.33 (in the middle) and during phase 0.59-0.61 (at the bottom) fitted with an absorbed power-law model.
Right column: The contour plot of $\chi^2$ between column density and photon index for each spectra. }
\label{inside}
\end{figure*}

\begin{table*}
 %\label{table}
\caption{ Spectral parameters for added Swift-XRT observations with quoted for 90\% confidence limits. }
\label{spectral_parameter}
  \centering
 \begin{tabular}{|c|c|c|c|c|c|}
 \hline
 &&&&&\\

Phase     		 &Column               & Photon          &Total			                      	&Upper limit of             &$\chi^2/d.o.f.$\\
        		 &Density              & Index           &Flux			                        &equivalent width of        &(without a gaussian \\
                 &N$_{H}$              &                 &(2.0-9.0 keV)		                	& Fe K$\alpha$ line         & component at\\ 
          	     &(10$^{22}$cm$^{-2}$) & ($\Gamma$)      &(10$^{-12}$ ergs.cm$^{-2}$sec$^{-1}$)	& (keV)                     &6.4 keV)\\
          
 \hline
                       &                              &                             &	            	&                       &\\
 0.91-1.06            &49$^{+17}_{-14}$      &1.9$^{+1.1}_{-0.9}$           &8.28$^{+0.41}_{-6.93}$     &0.346 keV              &0.729\\
                        &                              &                             &                  &                       &\\
 0.27-0.29            &50$^{+18}_{-14}$      &2.5$^{+1.3}_{-1.1}$           &18.10$^{+0.71}_{-10.03}$   &0.262 keV              &1.004\\
                        &                              &                             &                  &                       &\\
 0.55-0.57            &48$^{+16}_{-15}$      &2.3$^{+1.8}_{-1.3}$           &21.32$^{+0.17}_{-14.12}$   &0.345 keV              &0.529\\
                        &                              &                             &                  &                       &\\
    
 &&&&&\\
 \hline
 \end{tabular}
 \end{table*}
 
\begin{figure*}
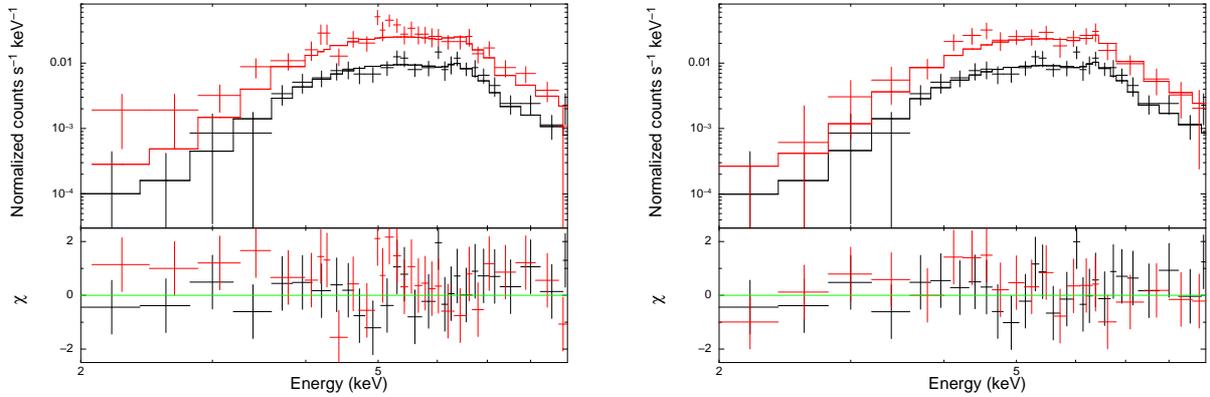

\centering
\includegraphics[scale=0.30, angle=-90]{simulataneous_28_8thMarch.eps} 
\hspace{5mm}
\includegraphics[scale=0.30, angle=-90]{simulataneous_56_8thMarch.eps}
\caption{The overlaid added X-ray spectrum during the low intensity and out of the low intensity state (left panel is with phase 0.31-0.33 and right panel is with phase 0.59-0.61 fitted with an absorbed power-law with gaussian model.}
\label{simultaneous}
\end{figure*}

\subsection{Orbital Intensity Profile Analysis}
We analyzed the most recent 15-50 keV {\it Swift}-BAT long term lightcurve and did a period search using `efsearch'. The time between the start and end time for the used BAT lightcurve is 5296 days and the effective total exposure is 45 Ms. We found a period of 366106 seconds ($\sim$ 4.24 days) with the epoch of MJD 53418.3 so that the minimum intensity is positioned at phase zero. We folded the lightcurve with the same period and epoch. We have used 32 time bins per period, same as in \cite{2014AAS...22332308C} and \cite{2015MNRAS.446.4148I} due to limited statistics in this source. The eclipse duration is for 3-4 bins.
\par
We have extracted the lightcurves with a default bin time of 2.5 seconds from each of the XRT observations and folded them with time period of 366106 seconds (i.e. $\sim$ 4.24 days) and with the epoch of MJD 53418.3.
%mentioned in \cite{2015MNRAS.446.4148I}, \textbf{\textit{It is better to determine the orbital profile again from the BAT light curve and use new period and epoch to fold the data as it will be more accurate? Mention both the period and the epoch.}}so that the minimum intensity is positioned at phase zero.\textbf{While folding we re-binned the lightcurve with 64 time bins}. \textbf{\textit{I suggest we give only 32 bins, same as in Corbet et al. and Islam et al. due to limited statistics in this source. The eclipse duration is for 3-4 bins. }}
These observations are almost evenly spread providing us full coverage of the entire orbital phase. The duration of the low intensity state is of about 1/5th of the total orbital period. The top and bottom panel of Figure \ref{BAT&XRT} shows the modulation in the orbital intensity profile from {\it Swift}-BAT and {\it Swift}-XRT observations respectively. 
The intensity fell to $\sim$ 30\% inside the eclipse phase indicating it to be partial in soft X-rays.
\par

%\textit{The previous sentence can be mentioned in the previous paragraph}.
Figure \ref{BAT&XRT} represents both the lightcurves from BAT (top panel) and XRT (bottom panel). The eclipse profile is narrow in both soft and hard X-rays. 
%The ingress and egress are not sharp, rather gradual.
The solid, dashed and dotted lines enclose the region of phase 0.91-1.09, 0.31-0.33 and 0.59-0.61 respectively which we have used for spectral analysis.  \\

\subsection{Spectral Analysis}
This partial eclipse profile, both in hard and soft X-rays, makes us inquire about the spectral nature of it. Individual XRT observations were too short for spectral analysis. We have therefore combined multiple observations in certain phase ranges for spectral analysis. There were 17 observations which are within the phase range of the partial eclipse (0.91 to 1.09). In Figure \ref{BAT&XRT}, these 17 observations fall in between the two solid lines. To make a comparison of the eclipse spectrum with spectrum during out of eclipse we have combined four out of eclipse observations in phase range 0.31-0.33 and four observations in phase range 0.59-0.61. Table \ref{table1} shows that both the phase range 0.31-0.33 and 0.59-0.61 have at least four observations, whereas there are fewer number of observations at other phases. Therefore, we choose these two particular phase to add up the spectra that can result into an improved statistic. The ranges of the out of eclipse spectra are also denoted in Figure \ref{BAT&XRT} with pair of dashed and dotted lines respectively. We have merged the ARF and RMF file separately for each observation. 
Then to get a particular added spectra, we have added the source and background spectra separately and added merged ARF-RMF files and combined them to get a background subtracted source spectra with RMF. The total exposure time for the added spectra during the partial eclipse phase is $\sim 15$ kilosec, and during phase 0.31-0.33 and 0.59-0.61, it is $\sim 4$ kilosec.\\

We have fitted all the three X-ray spectra at different orbital phases within the energy range 2.0-9.0 keV using XSPEC v12.9.1m. These spectra were modeled with a power law and photoelectric absorption due to the absorbing matter along our line of sight. The absorption column density, photon index and the total flux of these three spectra have been tabulated in Table \ref{spectral_parameter}. We do not notice any significant change in the column density inside and outside the eclipse. All of these spectra are in the energy range of 2.0-9.0 keV and are shown in Figure \ref{inside}. We have overlaid the eclipse spectrum with the out of eclipse spectrum at the two phase ranges 0.31-0.33 and 0.59-0.61
(Figure \ref{simultaneous}) and we see the spectral profiles are very similar inside and outside the low intensity state. Overall, the eclipse spectrum is lower by a factor of about 2-3.\\
For spectra with limited statistics, there is often dependence between the column density and photon index. To examine the dependence and to check the goodness-of-fit to the model, we have used the chi-square statistic. It provided the best fit value for column density and photon index for each of the three spectra. In the left column of Figure \ref{inside}, there are three $\chi^2$ contour plots. The ``+'' sign is where the $\chi^2$ is minimum, and $68\%$, $90\%$ and $99\%$ confidence contours are plotted with the solid, dashed and dotted lines. 
\par
An iron emission line at 6.4 KeV was detected in this source with the XMM EPIC-PN by \citet{Bodaghee2006}. Also, 6.4 keV lines are ubiquitous in HMXBs \citep{2015Gim}. However, in all the three spectra at different phases, we do not observe any significant presence of the iron emission line. For a possible presence of a line at 6.4 keV, we have estimated the upper limit of the equivalent widths at $90\%$ confidence level in Table \ref{spectral_parameter}.
 
%\begin{figure*}
%\centering
%\includegraphics[angle=-90,scale=0.35, trim={0 1.2cm 0 0}, clip=ture]{combine10_first5.eps}
%\hspace{5mm}
%\includegraphics[angle=-90,scale=0.35, trim={0 1.2cm 0 0}, clip=ture]{combine10_last5.eps}

%\caption{ Orbital intensity profile of 10 different HMXBs obtained using {\it Swift}-BAT long term lightcurves.
%In case of all of these
%HMXBs, the intensity in the eclipse goes to nearly zero.}
%\label{10HMXB}
%\end{figure*}

\begin{figure*}
\centering
\includegraphics[scale=0.37, trim={0 3.0cm 0 0.5cm},  angle=-90]{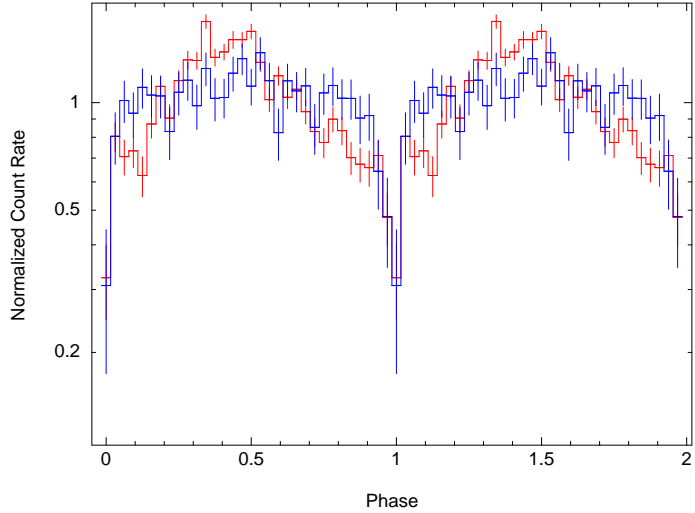}
\caption{The {\it Swift}-BAT long term lightcurves of 4U 0114+65 (red) and IGR J16393-4643 (blue) folded with the period of 11.596 days and 4.237 days respectively.}
\label{4U_IGR}
\end{figure*}

\section{Discussion}
We study the orbital intensity profile of the HMXB IGR J16393--4643 with {\it Swift}-XRT observations.
 The intensity dip during every orbital period observed in soft X-rays is similar to the intensity dip that has been seen in hard X-rays with {\it Swift}-BAT. The intensity falls off to $\sim 30$\% during the eclipse. \\

\begin{table*}
 %\label{table}
\caption{Ratio of Swift-BAT normalized count rate of inside and outside eclipse phase of the HMXBs reported in \citep{2015A&A...577A.130F}. }
\label{ratio}
  \centering
 \begin{tabular}{|c|c|c|c|}
 \hline
 &&&\\

Name of the source &	Normalized count rate (OOE$^a$)	& Normalized count rate (E$^b$)	&	Ratio (OOE/E)\\
\hline
&&&\\
Vela X-1 &		1.3	&	2.7e-2 &	48.15\\
&&&\\
LMC X-4	&		1.6	&	3.3e-2&	48.48\\
&&&\\
Cen X-3		&	1.3	&	2.5e-2&	52.00\\
&&&\\
4U 1700-377&		1.3&		3.4e-2&	38.24\\
&&&\\
4U 1538-522  & 		1.2&		6.9e-2&		17.39\\
&&&\\
SMC X-1	&		1.2&		6.5e-2	&	18.46\\
&&&\\
%SAX J1802.7-2017	(half of the counts have negative values)
%XTE J1855-026		(also negative counts)
EXO 1722-363&		1.2&		4.1e-2	&	29.26\\
&&&\\
OAO 1657-415&		1.4&		8.7e-2	&	16.09\\

&&&\\
 \hline
 \end{tabular}
 \begin{flushleft}
 $^a$ : Outside of eclipse phase excluding the eclipse ingress and egress\\
 $^b$ : Inside the eclipse phase\\
  \end{flushleft}
 
 \end{table*}

%\textcolor{red}{reference missing. Please check bibtex}
\citet{2010_Corbet} reported this significant modulation in the {\it Swift}-BAT lightcurve in every 4.24 days, and it was interpreted as an eclipse. This eclipse-like feature is not similar to the eclipse seen in other eclipsing HMXB systems, where the eclipse is usually broad, and the intensity during the eclipse falls to almost $\sim 1\%$ of that of the outside the eclipse. \citet{2015A&A...577A.130F} discussed 10 eclipsing HMXBs and presented the orbital intensity profile of the eclipsing HMXB sources like Cen X-3, 4U 1700-37, Vela X-1 etc. In Table \ref{ratio} we have tabulated the average normalized count rate for inside and outside the eclipse phase observed in 8 of those HMXBs from {\it{Swift}}-BAT lightcurve. 
We did not include XTE J1855-026 and EXO 1722-363, as the eclipse count rate measurement of these sources have large relative error. To make Table \ref{ratio}, we took the most recent long-term orbital lightcurve from {\it{Swift}}-BAT site and folded with the orbital periods and epoch times reported in \citep{2015A&A...577A.130F}. The eclipse profile of these HMXBs along with the ratio of the count rates makes it clearly evident that the partial eclipse seen in IGR J16393-4643 is narrower and much shallower. This casts doubt if the partial eclipse is really an eclipse or it is an X-ray dip taking place as a result of a periodic increase in absorption. From the {\it Swift}-XRT observations reported here we see that:\\
 
\par 
1) The photon index during the eclipse is comparable to the same outside eclipse. In other eclipsing X-ray binaries, the photon index is usually higher during eclipse due to energy dependence of the scattering. If at all, the photon index is smaller during the partial eclipse of IGR J16393-4643.\\

 \par 
 2) X-ray spectra of HMXBs during the eclipse are known to exhibit a substantial equivalent width of the iron line \citep{2019_Aftab}. It can be up to as much as 1.5 keV (Cen X-3: \citealt{1992ApJ...396..147N}, \citealt{2011ApJ...737...79N}; Vela X-1: \citealt{1994ApJ...436L...1N}). In spite of the limited statistics of the data we were able to put an upper limit of $\sim350$ eV to the equivalent width of the Fe-K$\alpha$ line present at 6.4 keV, which is considerably smaller than the iron line equivalent width of HMXBs in eclipse. \\
 
 \par
 3) There is no significant increase in the absorption column density during the low intensity state, so we can not infer whether the periodic decrease in intensity is a result of increased absorption in the stellar wind or not. 
 
 \par
In particular, this partial eclipse is very similar to the partial eclipse present in another HMXB 4U 0114+65 (a.k.a. 3A 0114+650), which
was also interpreted previously as an eclipse in literature with an orbital period of $\sim 11.59$ days \citep{1985_Crampton, 2006_Wen, 2007_Grundstrom}. In Figure \ref{4U_IGR},
We have presented the \textit{Swift}-BAT long-term lightcurve of 4U 0114+65 after folding it with 1001903.4 seconds ($\sim 11.596$ days) with an epoch of MJD 53421.5.
We can see the similarity in the intensity profile of IGR J16393-4643 and 4U 0114+65, that are plotted using {\it Swift}-BAT long-term observation of these two sources. However, spectral analysis of the low intensity feature of 4U 0114+65 with a {\it {Suzaku}} observation, gave a moderate value of absorption column density and a low value of the
equivalent width of Fe-K$\alpha$ line during the eclipse, which was in contrast with the high value of the equivalent width of Fe-K$\alpha$ 
line seen in other eclipsing HMXB spectra. These spectral characteristics along with detection of pulsations in low intensity state showed that this low intensity phase in 4U 0114+65 is not due to the eclipse, but it is an X-ray dip occurring due to the increased absorption in the stellar wind \citep{Pradhan2015}. Lower intensity during some narrow orbital phase range has also been reported in GX 301-2 \citep{2014Islam}. The low intensity phase seen in IGR J16393-4643 is most likely due to absorption in the stellar corona. A grazing eclipse by the companion star occurs due to absorption caused by the stellar corona. To estimate the inclination angles for grazing eclipse caused by the stellar corona, we follow the same method as in \citet{2015MNRAS.446.4148I} with an eclipse semi-angle of zero for nearly circular orbit and mass of the neutron star i.e. 1.4 M$_{\odot}$. The inclination angle is 65$^{\circ}$ - 77$^{\circ}$ for a main sequence B star with 10 R$_{\odot}$ and the range of 56$^{\circ}$ - 68$^{\circ}$ for a supergiant O9 star with 20 R$_{\odot}$. 
Further study of this low intensity phase with \textit{XMM-Newton} or \textit{Chandra} will be useful in studying the stellar winds in this system.

\section*{Acknowledgements}

% Entry for the table of contents, for this guide only
\addcontentsline{toc}{section}{Acknowledgements}
We thank the anonymous referee for helpful comments and suggestions. 
Authors present the study from the observations made by Neil Gehrels Swift Observatory, a mission by NASA. The data is
acquired from the High Energy Astrophysics Science Archive (HEASARC) Online Service provided by NASA/GSFC. Authors have also
used the public light-curves from {\it Swift}.

%%%%%%%%%%%%%%%%%%%% REFERENCES %%%%%%%%%%%%%%%%%%

\bibliography{bibtex}
\bibliographystyle{mnras}

%%%%%%%%%%%%%%%%%%%%%%%%%%%%%%%%%%%%%%%%%%%%%%%%%%

% Don't change these lines
\bsp	% typesetting comment
\label{lastpage}
\end{document}